\documentclass[conference]{IEEEtran}
\IEEEoverridecommandlockouts

\usepackage{array}
\usepackage{caption}
\usepackage{subcaption}
\usepackage{cite}
\usepackage{amsmath,amssymb,amsfonts}
\usepackage{algorithmic}
\usepackage{graphicx}
\usepackage{textcomp}
\usepackage{xcolor}
\usepackage{tablefootnote}
\usepackage{float}
\usepackage{ulem}
\def\BibTeX{{\rm B\kern-.05em{\sc i\kern-.025em b}\kern-.08em
    T\kern-.1667em\lower.7ex\hbox{E}\kern-.125emX}}

\renewcommand{\baselinestretch}{0.9}  
\begin{document}

\newcommand{\blue}[1]{\textcolor{blue}{#1}}
\newcommand{\red}[1]{{\textcolor{red}{#1}}}
\newcommand{\green}[1]{\textcolor{green}{#1}}
\newcommand{\magenta}[1]{\textcolor{magenta}{#1}}
\newcommand{\cyan}[1]{\textcolor{cyan}{#1}}
\newcommand{\fa}[1]{\textcolor{magenta}{Fatemeh: #1}}
\newcommand{\mg}[1]{\textcolor{orange}{Mohammed: #1}}
\newcommand{\bh}[1]{\textcolor{purple}{#1}}
\newcommand{\yellow}[1]{\colorbox{yellow}{#1}}
\newcommand{\ak}[1]{\textcolor{blue}{Andre: #1}}

\title{5G Wings: Investigating 5G-Connected Drones Performance in Non-Urban Areas\\
\thanks{{This material is based upon work supported by T-Mobile, the Air Force Office of Scientific Research under award number FA9550-20-1-0090, and the National Science Foundation under Grant Numbers  CNS-2120485 and CNS-2232048.}}
}

\author{\IEEEauthorblockN{	Mohammed Gharib, Bryce Hopkins, Jackson Murrin, Andre Koka, Fatemeh Afghah. 
% 	\thanks{Distribution A: Approved for Public Release, distribution unlimited.  Case Number 88ABW-2019-???? on ?? Apr. 2019. The work of F. Afghah, J. Ashdown and K. Turck was supported by AFRL. }}
	}
	\IEEEauthorblockA{Department of Electrical and Computer Engineering, Clemson University, Clemson, SC,  USA.\\ E-mail:\{alghari, bryceh, jmurrin, arkoka, fafghah\}@clemson.edu} }

\maketitle

\begin{abstract}
Unmanned aerial vehicles (UAVs)  have become extremely popular for both military and civilian applications due to their ease of deployment, cost-effectiveness, high maneuverability, and availability. Both applications, however, need reliable communication for command and control (C2) and/or data transmission. Utilizing commercial cellular networks for drone communication can enable beyond visual line of sight (BVLOS) operation, high data rate transmission, and secure communication. However, deployment of cellular-connected drones over commercial LTE/5G networks still presents various challenges such as sparse coverage outside urban areas, and interference caused to the network as the UAV is visible to many towers. Commercial 5G networks can offer various features for aerial user equipment (UE) far beyond what LTE could provide by taking advantage of mmWave, flexible numerology, slicing, and the capability of applying AI-based solutions.
Limited experimental data is available to investigate the operation of aerial UEs over current, without any modification, commercial 5G networks, particularly in suburban and NON-URBAN areas. In this paper, we perform a comprehensive study of drone communications over the existing low-band and mid-band 5G networks in a suburban area for different velocities and elevations, comparing the performance against that of LTE. It is important to acknowledge that the network examined in this research is primarily designed and optimized to meet the requirements of terrestrial users, and may not adequately address the needs of aerial users. 
This paper not only reports the Key Performance Indicators (KPIs) compared  among all combinations of the test cases but also provides recommendations for aerial users to enhance their communication quality by controlling their trajectory. 
\end{abstract}

\section{Introduction}
\label{sec:intro}

Unmanned aerial vehicles, also known as drones, have been attracting significant attention from industry, academia, and the military by being cost-effective, available, versatile, and having high maneuverability. Various applications such as entertainment, package delivery, border surveillance, and remote sensing are a few examples to name \cite{ALSAMHI2021102505}. Regardless of the intended application of the aerial vehicle, reliable C2 communication is crucial in all application domains to ensure safe operation. Additionally, high data rate communication is of significant importance in UAVs' applications in disaster relief, augmented reality, and environmental and infrastructure monitoring\cite{ROVIRASUGRANES2022102790}. BVLOS operation, in turn, paves the way for various new applications of autonomous UAVs. However, it needs an ultra-reliable low latency communication \cite{kthBvlos}. 
Technologies such as WiFi \cite{wifiUav}, LoRA \cite{loraUav}, and WiMAX \cite{wimaxUav} are available for UAV communication. However, the short communication range of WiFi, the low bandwidth of LoRA, 
and the latency, as well as the lack of support for highly mobile users in WiMAX, are the limiting factors that make the utilization of aforementioned technologies more concerning. More importantly, these technologies do not often meet the security expectations for UAV communication. 
The wide deployment, availability, high data transmission rates, reliability, and security of cellular networks make them ideal candidates for UAV communication.  
However, cellular networks are essentially designed and optimized to serve terrestrial users and might not be ready to serve aerial users, without modifications. Besides 3D coverage gaps due to cellular antennas being tilted toward the terrestrial  users,  frequent handovers, interference caused by UAVs to neighbor towers, mobility among different providers to maintain connectivity, and UAV identification are among the challenges of serving drones by cellular networks.  

5G features such as low latency, higher bandwidth, ultra-reliability, strong security, highly directional signal delivery enabled by massive MIMO, slicing, mmWave, flexible numerology, and advanced beamforming techniques, as well as the capability of applying AI-based solutions such as AI-based path planning for interference mitigation and coverage maximization put 5G in a unique position to serve aerial users. 
Current T-mobile 5G deployment includes stand-alone (SA) and non-standalone (NSA) networks and provides layer cake spectrum strategy to supply coverage in three different bands, low-band which covers frequencies less than 1 GHz, mid-band which covers 1 to 6 GHz, and high-band, also referred to as mmWave, which covers frequencies greater than 24 GHz. The latter provides much higher bandwidth compared to the mid-band 5G, and hence drastically increased throughput. However, to face the frequency-dependent path loss and to provide sufficient coverage for its high-frequency signal, it requires directional antennas in both UE and basestation (BS) side \cite{zhangInfocom}.

Some challenges regarding 5G-connected drones  are the coverage at different elevations, the performance at higher elevations, and the interference of the aerial UE on the primary terrestrial users. While the latter category is out of the scope of this paper, the concern with the first two categories rises from the fact that the tilt angle of cellular antennas is optimized such that, considering  on-the-ground obstacles,  signal coverage is maximized and interference is minimized for terrestrial users. The former studies of cellular network communication for aerial users (discussed in Section (\ref{sec:relatedWork}), showed evidence of cellular coverage for aerial vehicles flying in low altitudes allowed by FCC/FAA. While there are valuable works in the literature \cite{5G_end2end,5gTracker,ChicagoMiami}, the problem of cellular coverage has not yet been sufficiently studied for commercial 5G networks. Most of the literature considered the evaluation of 5G networks for terrestrial users or utilized a testbed with a private 5G network for the measurement. An exhaustive performance analysis is required to assess the feasibility of using commercial 5G networks in different geographical areas and network conditions. 

In this paper, we analyze the coverage and the performance of the cellular 5G network for aerial users via exhaustive real measurements over a commercial LTE/5G network. We perform a comprehensive study on the low-band 5G, mid-band 5G, and LTE networks by measuring  KPIs at different elevations for an aerial user with different velocities  in a sub-urban area near Clemson University, Clemson, SC USA.
The aim is to investigate several open research questions about cellular-connected UAV communication including but not limited to the variation of different network KPIs at different deployment modes, the impact of elevation and velocity on the network's KPIs, and the 3D coverage in a sparse suburban area compared to the 2D ground-level coverage. 
To investigate the impact of deployment modes on the KPIs, we equipped a DJI M30T drone with three cellphones, each running RFInsight, a T-mobile application to measure cellular network performance. RFInsights provided the collected data as plain text files, which were then cleaned to remove  superfluous data collected during takeoff, landing, and times other than the main flights. The cleaned logfiles were then analyzed to extract beneficial, meaningful information. The details of data gathering, cleanup, and processing are reported in Section (\ref{sec:testProcedure}) and the results from the processed data are represented in Section (\ref{sec:results}). To investigate the aforementioned questions on the capability of the current, without any modification, LTE/5G networks to serve low-altitude drones in suburban areas,  Reference Signal Received Power (RSRP), Reference Signal Received Quality (RSRQ), Reference Signal Signal to Noise Ratio (RSSNR), uplink throughput, and downlink throughput are measured and analyzed for all the combinations of the tested velocities  and elevations, for LTE, low-band 5G, and mid-band 5G networks. 

The main contribution of this paper is to exhaustively evaluate the performance of the current 5G cellular network in its low and mid bands for aerial communication and compare it with that of LTE in sub-urban areas. The additional contribution of this paper is to offer trajectory planning recommendations for aerial users which intend to utilize commercial 5G as their communication platform, to enhance their communication performance. The rest of this paper is organized as follows. In Section (\ref{sec:relatedWork}), we review the state-of-the-art with an exhaustive comparison among the related works. In Section (\ref{sec:testProcedure}), we present the test, data collection, and data pre-processing. We show the results in Section (\ref{sec:results}). Finally, we conclude the paper and present future directions in Section (\ref{sec:conclusion}).
\section{Related Work}
\label{sec:relatedWork}

\newcounter{mpFootnoteValueSaver}
\begin{center}
\begin{table*}[!h]
\caption{An Overview of Related Works } 
\label{t_compare}
\resizebox{\textwidth}{!}{
\setcounter{mpFootnoteValueSaver}{\value{footnote}}
\begin{tabular}{|c|c|c|c|c|c|c|c|c|c|c|c|c|c|}
\hline
Related Work & UAV& LTE & 5G & RSRP & SINR & RSRQ&Speed&Elev.&\multicolumn{2}{c|}{Throughput} & C/T\footnotemark&Area\footnotemark&Measurement Tool\\\cline{10-11}%&Other measurements 
&&&&&&&&&UL&DL&&&\\\hline
Narayanan et al.\cite{variegatedLook}&$\times$ & \checkmark & \checkmark &  \checkmark & \checkmark & \checkmark&\checkmark&$\times$ &\checkmark & \checkmark &C& U &Ookla\\%&Power usage \\
Rochman et al.\cite{thermal} &$\times$ & \checkmark & \checkmark &  \checkmark & \checkmark & \checkmark &$\times$&$\times$& \checkmark & \checkmark &C& U &SigCap,NSG\footnotemark\\%&UE skin temperature \\
Narayanan et al.\cite{5gTracker}&$\times$&$\times$& \checkmark  & \checkmark & \checkmark & \checkmark &$\times$&$\times$& \checkmark &\checkmark&C&U&5GTracker\\%&Cell phone information \\
Rochman et al. \cite{ChicagoMiami}&$\times$&\checkmark & \checkmark &  \checkmark &$\times$ & \checkmark &$\times$&$\times$&$\times$& \checkmark&C&U&SigCap, FCCS\footnotemark, NSG\\%&Distance and Cell ID \\
Maeng et al.\cite{AERIQ}&\checkmark&\checkmark&$\times$ & \checkmark &$\times$ & \checkmark &\checkmark&\checkmark& $\times$&$\times$&T&R&KNO\footnotemark, QualiPoc, TEMS \\%&PSS\footnote{Primary Signal Synchronization} \\
Gharib et al.\cite{Exhaustive}&\checkmark &\checkmark & $\times$& \checkmark & \checkmark & \checkmark&\checkmark&\checkmark& \checkmark&\checkmark&C&R&TEMS \\%& Handover Analysis\\
Lin et al.\cite{SimulationsDesign}&\checkmark &\checkmark &$\times$ & \checkmark & \checkmark &$\times$&\checkmark&\checkmark& \checkmark&\checkmark&C&S&TEMS \\%&RS-SINR, Latency \\
Amorim et al.\cite{RadioModeling}&\checkmark&\checkmark&$\times$ &  \checkmark & \checkmark &$\times$ &\checkmark&\checkmark&$\times$&$\times$&C&R &RF Antenna Scanner\\%&SINR degradation \\
Amorim et al.\cite{DualNetworkControlLink}&\checkmark&\checkmark& $\times$& \checkmark &$\times$ & \checkmark &\checkmark&\checkmark&\checkmark &\checkmark&C&U&Dedicated App\\%&DL-MCS \\
Sekander et al.\cite{Multi-tier}&\checkmark&$\times$&\checkmark &$\times$ & \checkmark &$\times$ &$\times$&$\times$&$\checkmark$&\checkmark&T&U, S, R&Simulation\\%&Multiple drone use \\
Festag et al.\cite{5G_end2end}&\checkmark&\checkmark&\checkmark &  \checkmark & \checkmark &$\times$&\checkmark&\checkmark &\checkmark&\checkmark&T&S &ICPM echo, iPerf\\%&Long-distance flights \\
Säe et al.\cite{5GMeasure_UAV}&\checkmark&\checkmark& \checkmark&  \checkmark & $\times$& \checkmark &\checkmark&\checkmark&\checkmark&\checkmark&T&S &MediaTek, QualiPoc\\%&SS-RSRP and SS-RSRQ \\
Raouf et al.\cite{raouf2023spectrum}&\checkmark&\checkmark& \checkmark &\checkmark &$\times$ &$\times$&$\times$&\checkmark& \checkmark &\checkmark &C&U, R &USRP \& Py Script\\%&Frequency Spectrums \\
\textbf{This Work}&\checkmark&\checkmark&\checkmark&\checkmark&\checkmark&\checkmark&\checkmark&\checkmark&\checkmark&\checkmark&C&S&RFInsights\\%&Altitude and Distance\\
\hline
\end{tabular}}
\end{table*}
\end{center}
As commercial 5G stand-alone (SA) and non-stand-alone (NSA) networks have been deployed by commercial phone carriers, there has been significant interest in using these networks for UAV communication. To discover the feasibility of this, several measurement-based studies of current commercial 5G capabilities have been performed. Generally, we categorize the related work of this paper into three categories, general 5G measurement studies over commercial networks, LTE UAV measurement studies, and 5G UAV measurement studies. The first category includes measurement-based studies that focus exclusively on terrestrial users \cite{zhangInfocom,thermal,5gTracker,ChicagoMiami}. While these studies do not consider UAV communication, they are valuable for understanding the variations of different 5G  KPIs. %The other two categories focus on UAV cellular-connected communication and hence, they are the interest of this paper. 

\textbf{5G Performance Analysis for Terrestrial Users}:
 Narayanan et al. \cite{variegatedLook} analyzed cellular throughput, handover, and energy performance of deployed 5G networks. The authors concluded that mmWave 5G networks always have higher performance than 4G networks when line-of-sight connection is guaranteed, but leads to much higher energy consumption for communications with lower data transfer requirements \cite{variegatedLook}. In \cite{thermal}, Rochman et al. conducts one of the first studies of sustained 5G mmWave coverage across multiple United States cities, performing tests in Chicago, Miami, and San Francisco. They found that sustained mmWave throughput is limited by the temperature of the UE, which rises as intensive mmWave use continues. 
Analysis of the standard KPIs for 5G has proven to be difficult, as cellular providers are not transparent with their current 5G deployments. To combat this, Narayanan et al. \cite{5gTracker} present an open-source tool that allows for crowd-sourcing of large 5G KPI datasets using android smartphones. Other smartphone measurement tools are used by Rochman et al. in \cite{ChicagoMiami}, to measure 4G and 5G performance. In \cite{ChicagoMiami}, authors studied 5G deployments across multiple cities and found that 5G networks of all types are highly sensitive to physical obstructions compared to 4G networks. 

\noindent\textbf{4G/LTE Performance for UAV Communication}:
In \cite{AERIQ}, an analysis of air-to-ground link propagation characteristics is performed  using a tool introduced in the same work, referred to as AERIQ.
This paper extracts RSRP information from a variety of experiments and proves the testing capabilities of their AERIQ platform, while also finding that an increase in 3D distances results in a decrease in coherence bandwidth, due to the effects of ground reflection being more pronounced at longer distances. Gharib et al. \cite{Exhaustive} performed an analysis in rural areas, looking at communication characteristics between an aerial UAV and a commercial LTE network in Flagstaff, AZ. They concluded that rural areas experience better signal quality and fewer handover processes for low-elevation flights, and the increase in UAV velocity causes a slight degradation in the performance of the 4G network. 

Xingqin et al. conducted testing in a suburban area for different elevations, using the TEMS Pocket smartphone application to take measurements. They found that 4G networks could support the initial deployment of drones, but infrastructure improvements would be required to support drones that operated at higher altitudes and speeds. This conclusion is supported by Amorim et al. \cite{RadioModeling}, where they performed flight tests to find a mathematical model for path loss of radio communication among multiple UAVs. Using a scanner mounted on a UAV, they recorded RSRP and distance information at several points in 3D space to calculate a path loss sample at each point. They found that path loss attenuation of radio signals is significantly higher at high altitudes compared to ground levels. 
Amorim et al. \cite{DualNetworkControlLink} analyzed the reliability of the UAV control link. After flying at 15-, 40-, and 100-meter elevations and measuring latency, they concluded that the control link connection through LTE-A cellular networks falls short of the 99.9\% reliability goal defined by 3GPP. In \cite{DualNetworkControlLink}, authors proposed the use of a dual-operator scheme, which manages to reach the desired performance on LTE networks in most cases. 

\noindent\textbf{5G Network for UAV Communication}: 
Bor-Yaliniz et al. \cite{is5Gready} performed an analysis to determine if commercial 5G networks should be used to support drone use. They found that there is no simple way to determine how best to utilize UAVs with 5G networks. However, by comparing a large number of 5G design options, like relaying and cloud-RAN, they concluded that 5G slicing and modularity are network elements that provide flexibility to be utilized to support drones on the networks. They, however, performed no flight testing of 5G networks. An early instance of 5G UAV flight testing is provided by Festaget al. \cite{5G_end2end}. They performed long-distance flight testing over a 7km course at a height of 100m, analyzing end-to-end latency. They concluded that cellular communication between UAVs and 5G base stations meets the demands for video streaming in principle, even when the network has not been optimized for aerial users. They also state that network slicing abilities in 5G networks have the potential to improve network performance further. 

Säe et al. \cite{5GMeasure_UAV} considered an alternative to traditional drive testing, by conducting flight tests using an established, commercial-grade 5G test network. They were able to calculate 4G and 5G radiation patterns within the 3D space. This led them to conclude that UAV flight testing with a smartphone as the measurement device is a reasonable alternative to drive testing for the purposes of evaluating the coverage capabilities of a given 5G network. Raouf et al. \cite{raouf2023spectrum} performs network power measurements using UAV flight testing, allowing for comparison of aerial coverage between 4G LTE and commercial 5G low-mid band networks. Using the NSF AERPAW measurement platform in Raleigh NC,  Raouf et al. performed flights at 140m and 180m for the urban and rural sites, respectively. They found that generally the measured signal power increases as UAV altitude increases, due to the better line-of-sight characteristics at high altitudes. They also found that the spectrum of down-link frequencies is significantly more crowded than the up-link spectrum in both environments. 
Table (\ref{t_compare}) summarizes a comprehensive comparison among the literary works and compares them with our work. 
Overall, there is a continued need to investigate deployed 5G potential for aerial communications. Previous works in this area have performed a speculative analysis of drone use with 4G and 5G networks, while others have performed physical flight testing. This paper provides a comprehensive comparison between UAV altitude, velocity, and the connected cellular network band. In addition, this paper includes altitude and velocity testing of a real 5G network, as opposed to a 5G testbed.
\stepcounter{mpFootnoteValueSaver}%
    \footnotetext[\value{mpFootnoteValueSaver}]{%
      Commercial network versus testbed measurement.}
  \stepcounter{mpFootnoteValueSaver}
    \footnotetext[\value{mpFootnoteValueSaver}]{ R: Rural, S: Suburbun, U: Urbun     }
    \stepcounter{mpFootnoteValueSaver}
    \footnotetext[\value{mpFootnoteValueSaver}]{Network Signal Guru      }
    \stepcounter{mpFootnoteValueSaver}
    \footnotetext[\value{mpFootnoteValueSaver}]{FCC Speedtest      }
    \stepcounter{mpFootnoteValueSaver}
    \footnotetext[\value{mpFootnoteValueSaver}]{Keysight NEMO Outdoor}

\section{Data Collection and Processing}
\label{sec:testProcedure}

\begin{figure*}
         \begin{subfigure}{0.195\textwidth}
         \centering         \includegraphics[width=\textwidth, height=2.37cm]{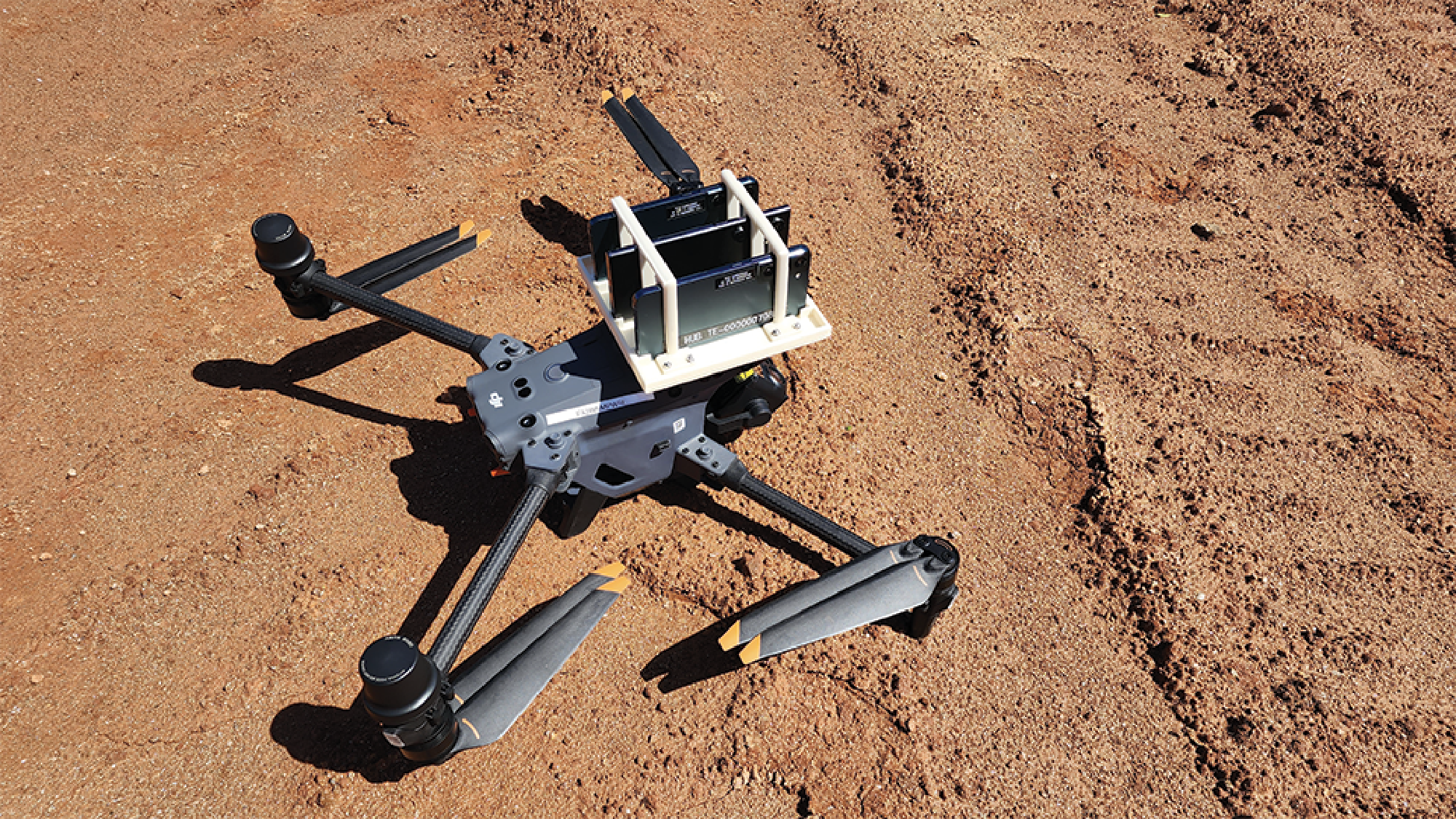}
         \caption{Drone Setup}
         \label{fig:drone}
     \end{subfigure}
     \begin{subfigure}{0.195\textwidth}
         \centering
         \includegraphics[width=\textwidth, height=2.37cm]{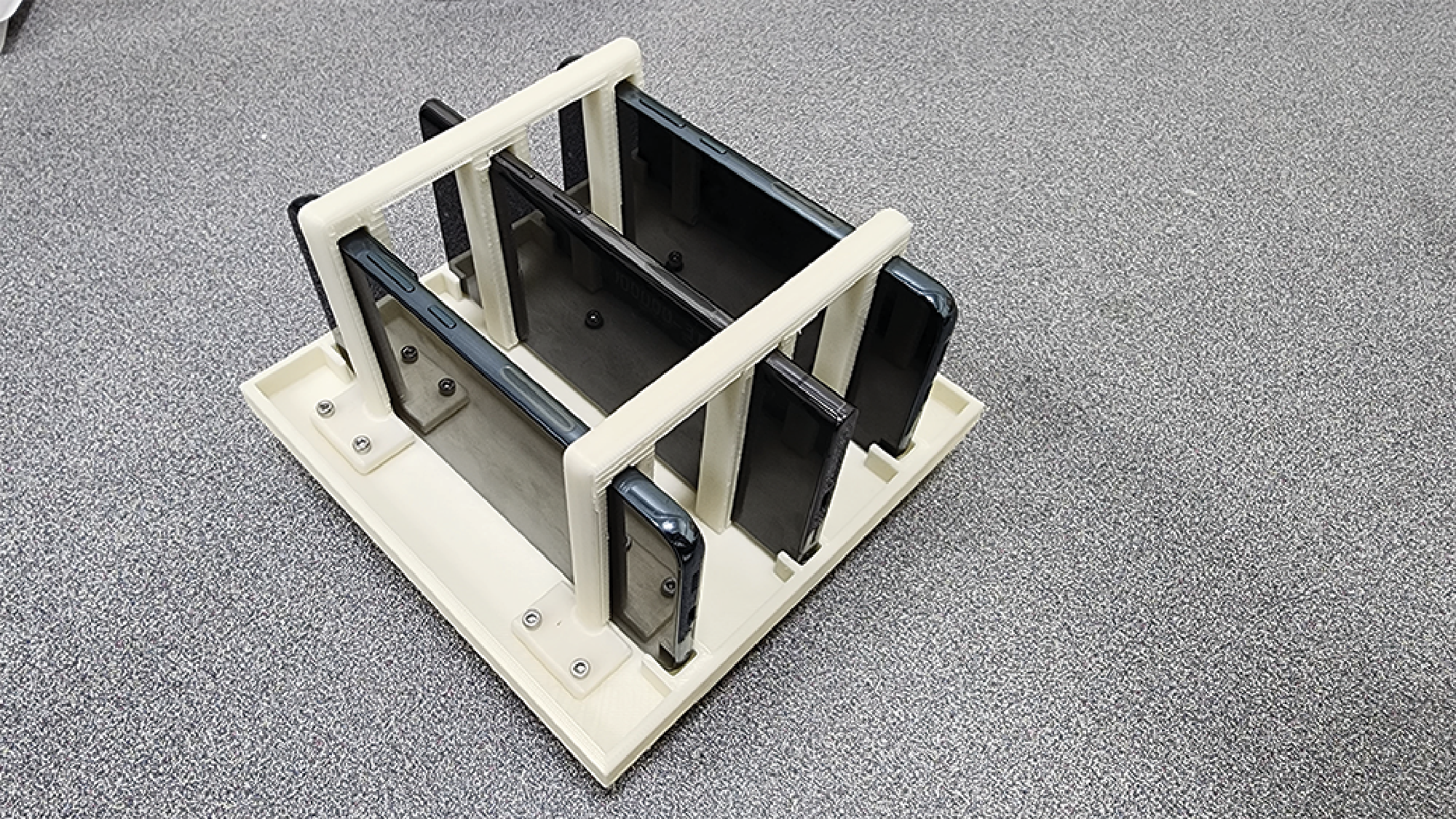}
         \caption{Phone Setup}
         \label{fig:phones}
     \end{subfigure}
     \begin{subfigure}{0.195\textwidth}
         \centering
         \includegraphics[width=\textwidth, height=2.37cm]{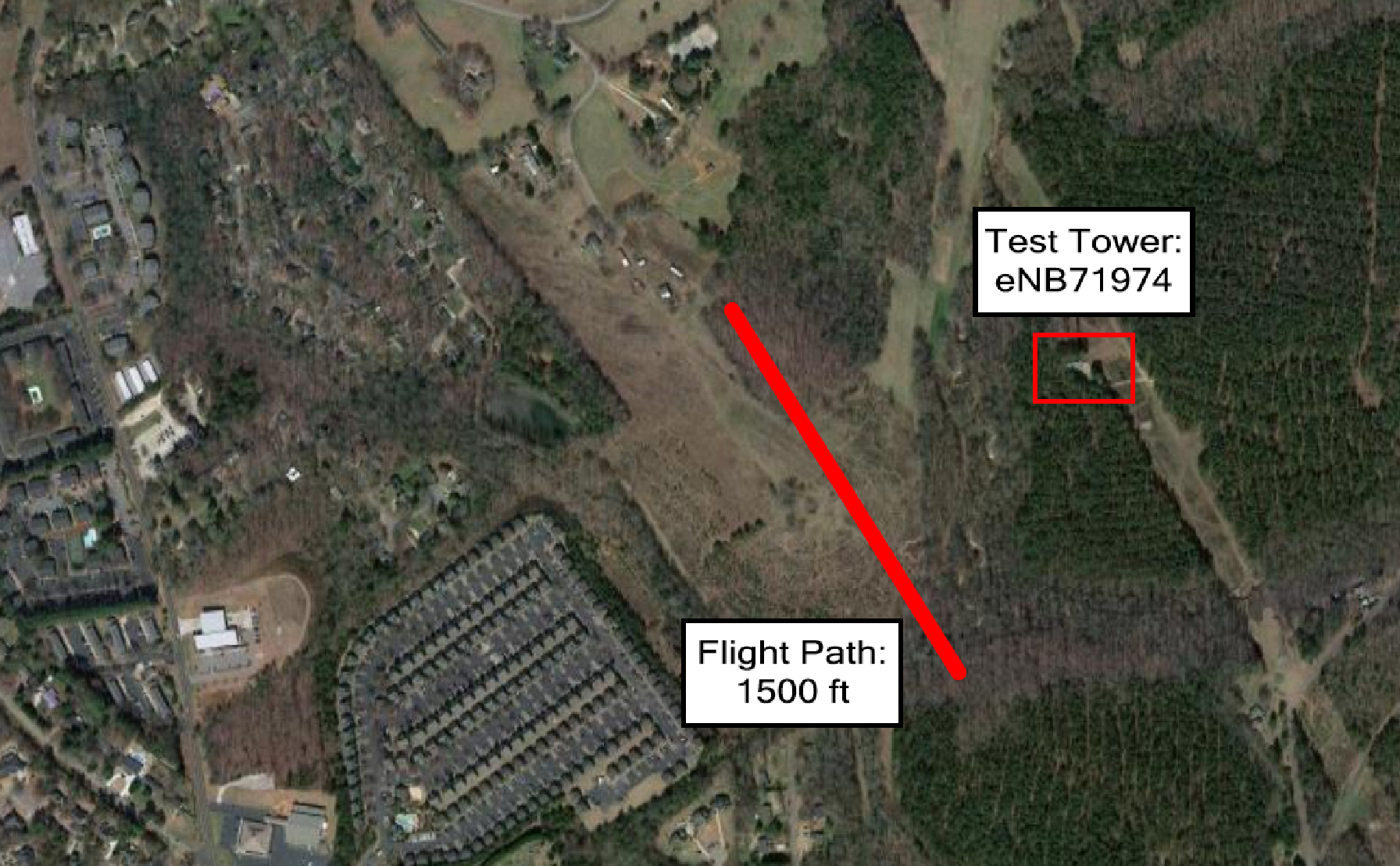}
         \caption{Flight Test Map}
         \label{fig:test_map}
     \end{subfigure}
     \begin{subfigure}{0.195\textwidth}
         \centering
         \includegraphics[width=\textwidth]{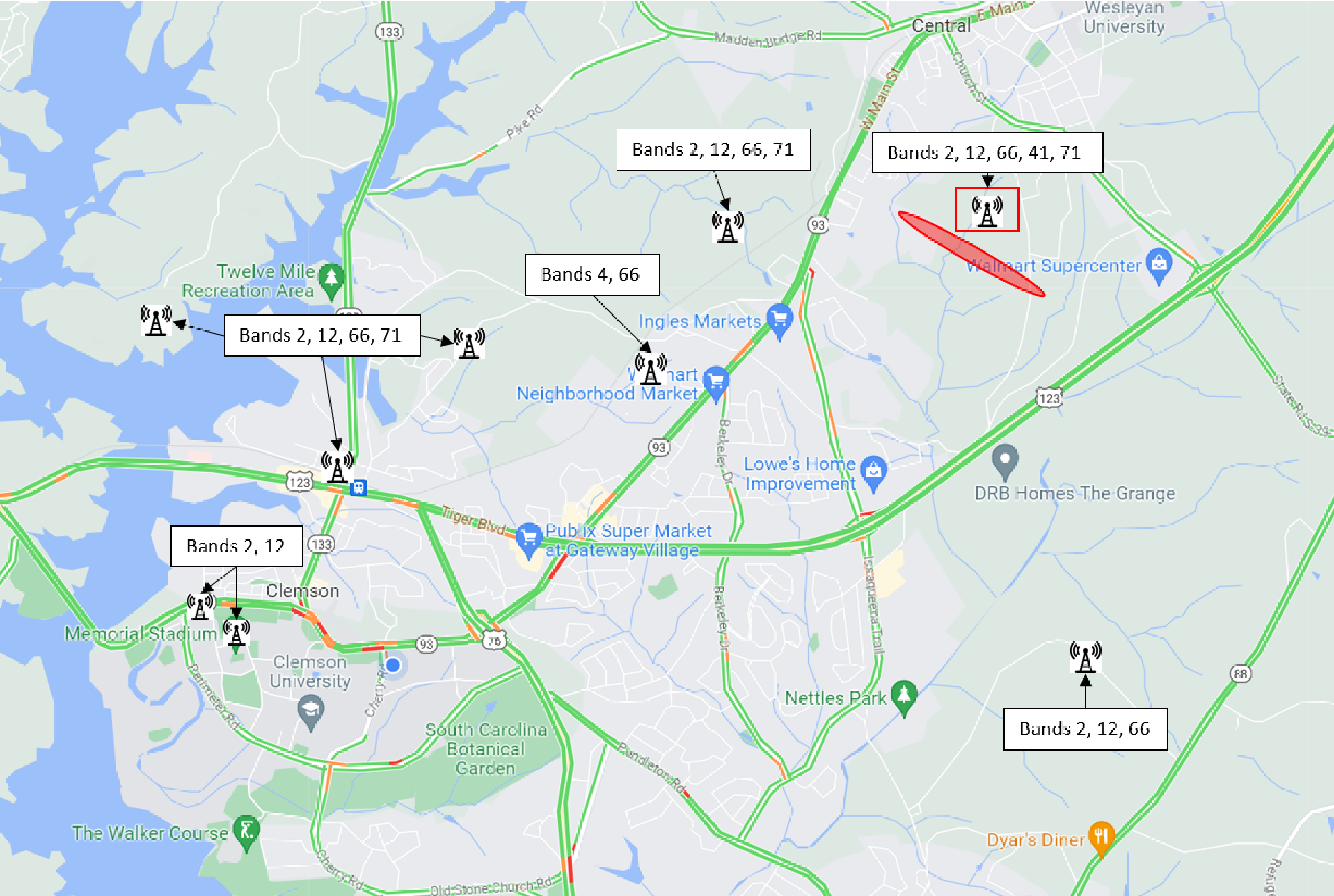}
         \caption{Overall View}
         \label{fig:overall_view}
     \end{subfigure}
     \begin{subfigure}{0.195\textwidth}
         \centering
         \includegraphics[width=\textwidth]{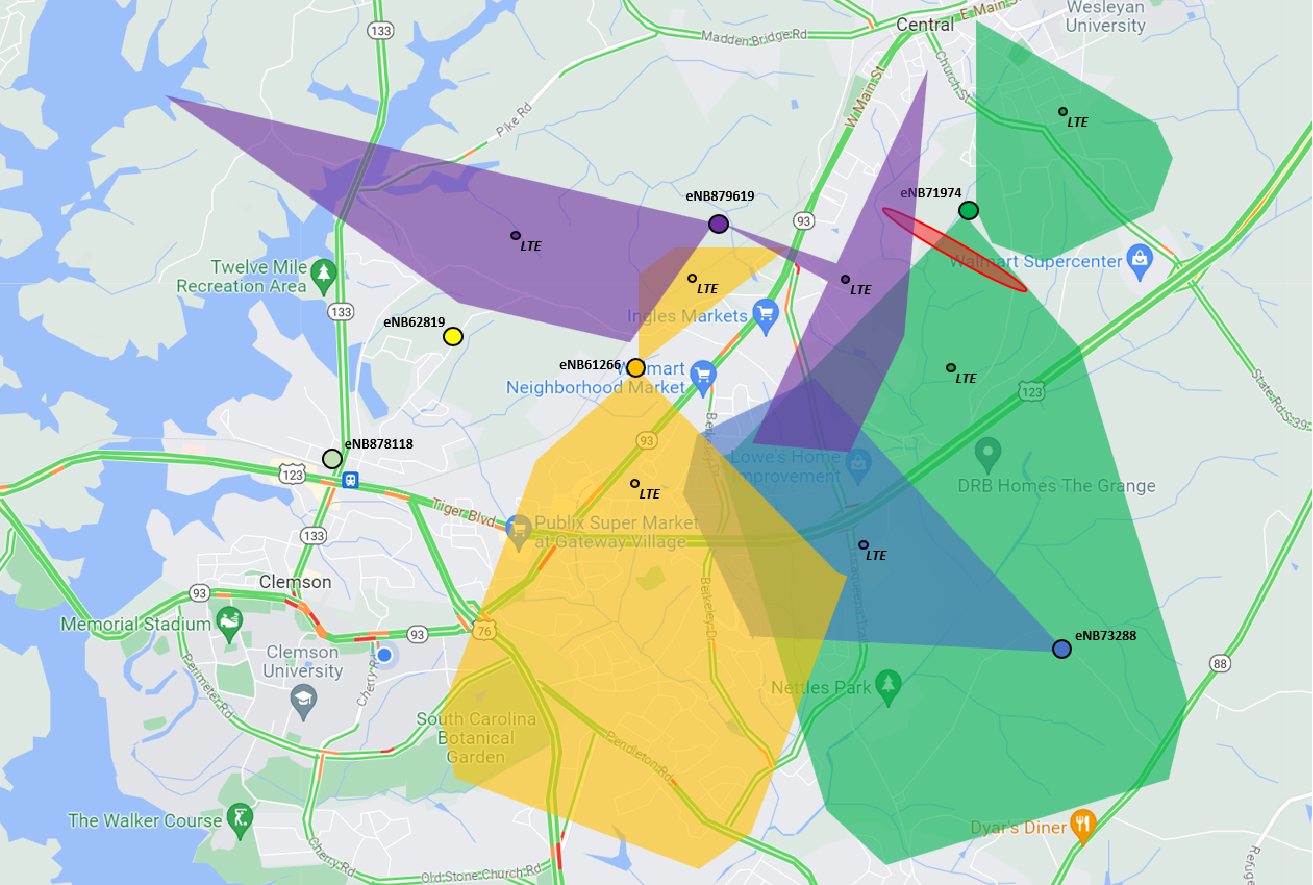}
         \caption{Nearby Cell Coverage}
         \label{fig:geometric_view}
     \end{subfigure}
        \caption{Measurement Setting}
        \label{fig:three graphs}
\end{figure*}

\begin{figure}
         \begin{subfigure}{0.45\columnwidth}
         \centering         \includegraphics[width=\textwidth,height=4.3cm]{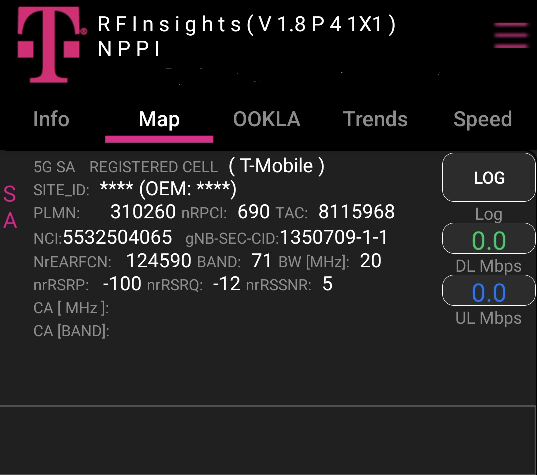}
         \caption{}
         \label{fig:rfi1}
     \end{subfigure}
     \begin{subfigure}{0.45\columnwidth}
         \centering
         \includegraphics[width=\textwidth,height=4.3cm]{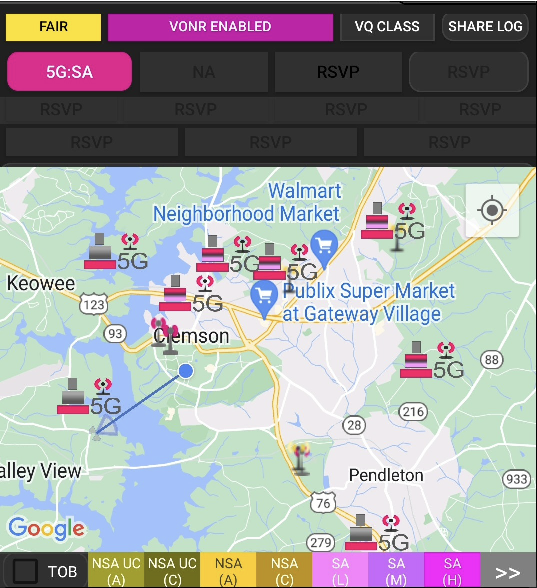}
         \caption{}
         \label{fig:phones}
     \end{subfigure}
             \caption{RFInsight Application Snapshot }
        \label{fig:rfi}
\end{figure}

In this section, we describe the procedure of data collection, as well as the data log file process. Our physical testing equipment is shown in Fig. (\ref{fig:drone}). It was comprised of a commercial DJI Matrice 30T drone, two Samsung S22+, and one Samsung S22 Ultra cellphones, and a 3D printed mount to securely attach the cellphones on top of the drone. Fig. (\ref{fig:phones}) shows the phones and the printed mount together before the installation on the drone. Our data log files were collected with T-Mobile's in-house RFInsights application, represented in Fig. (\ref{fig:rfi}), with built-in Ookla SpeedTest connected to different servers for uplink and downlink throughput measurement. RFInsight collects the data in a time granularity of one second. Our data was collected at an active construction zone in Central, South Carolina as seen in Fig. (\ref{fig:geometric_view}). This site had been deforested and leveled leaving an easy area to fly a drone in with minimal obstacles. Central, South Carolina is a suburban town next to the city of Clemson, South Carolina. All of the tests were conducted on the T-Mobile cellular network. Fig. (\ref{fig:overall_view}) shows an overall view of the city with the T-Mobile base stations, and the test area colored by red. Throughout the testing process, we detected signals from four base stations. Fig. (\ref{fig:geometric_view}) shows those BSs and the corresponding on-ground coverage areas, where the test area is highlighted in red. 
It is worth mentioning that the main practical concern in performing cellular network measurements with regular cellphones is  overheating \cite{thermal}.
As touched on previously, the est setup utilized  three different cell phones at once on the drone, exacerbating the overheating concern. To prevent overheating, a special 3D printed mount was designed that attaches to the top of the UAV, firmly securing the three cell phones. The  open-air design allowed for the phones to be naturally cooled. Once the 3D printed mount was employed the overheating problem was eliminated, even when running Ookla SpeedTest which produced the most heat of any testing conducted. The full setup can be seen in Fig. (\ref{fig:drone}). 

With the ability to collect three different data sets at the same time, we compared LTE, 5G low-band, and 5G mid-band as mmWave towers were not available in this suburban test site. The phones were able to be locked on specific T-Mobile bands exclusively. The first S22+ was set to all LTE bands, later being seen that the phone was connected on band 66 the whole flight. Band 66 is the 10MHz channel bandwidth on 2.175 GHz for downlink and 1.775 GHz for uplink. The second S22+ was set to test 5G mid-band on band n41 at 2.5 GHz frequency with 100 MHz channel bandwidth. This left the S22 ultra, set to test 5G low-band on band n71 at 600 MHz frequency with 15 MHz channel bandwidth. Nine different test scenarios were investigated, each with a specified elevation and velocity.  The different elevations flown at in the test scenarios are 400 feet, 300 feet, and 200 feet. For each of these elevations, three different velocities were tested: 30 mph, 20 mph, and 10 mph. This led to having nine different data scenarios to analyze the state of communication at different altitudes and velocities. For each test scenario, the drone was flown in a 1,500-foot linear path, illustrated in Fig. (\ref{fig:test_map}), stopped at the far end, and then flown back 1,500 feet to the starting point. The log files are then collected and pre-processed to remove unnecessary data such as those belonging to the time before achieving exact elevation or speed.

\section{Measurement Results for 5G-Connected UAV }
\label{sec:results}
\begin{figure*}
         \begin{subfigure}{0.333\textwidth}
         \centering         \includegraphics[width=\textwidth]{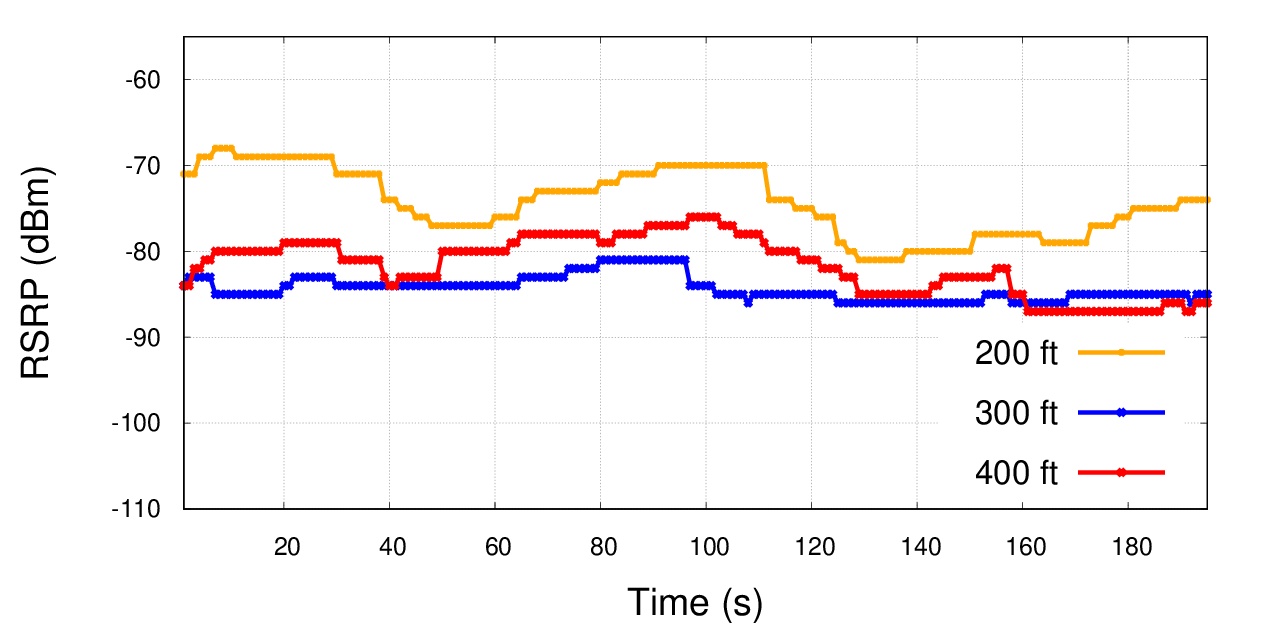}
         \caption{LTE}
         \label{fig:rsrpLte}
     \end{subfigure}
     \begin{subfigure}{0.333\textwidth}
         \centering         \includegraphics[width=\textwidth]{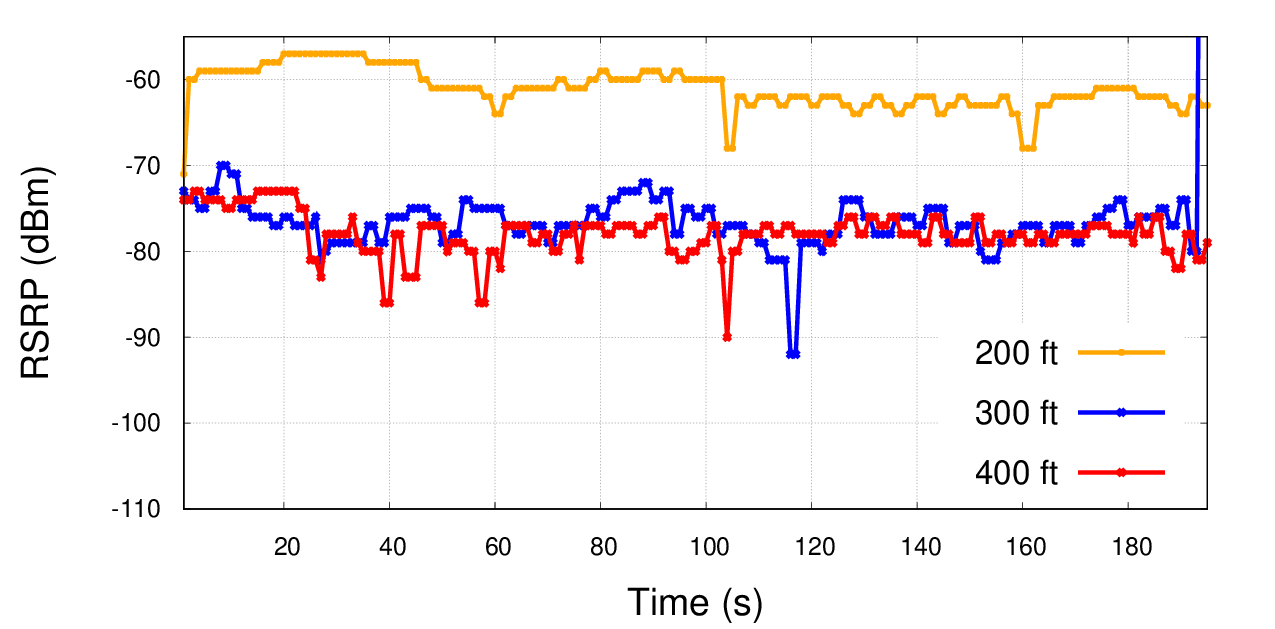}
         \caption{Low-band 5G}
         \label{fig:rsrpLow}
     \end{subfigure}
     \begin{subfigure}{0.333\textwidth}
         \centering         \includegraphics[width=\textwidth]{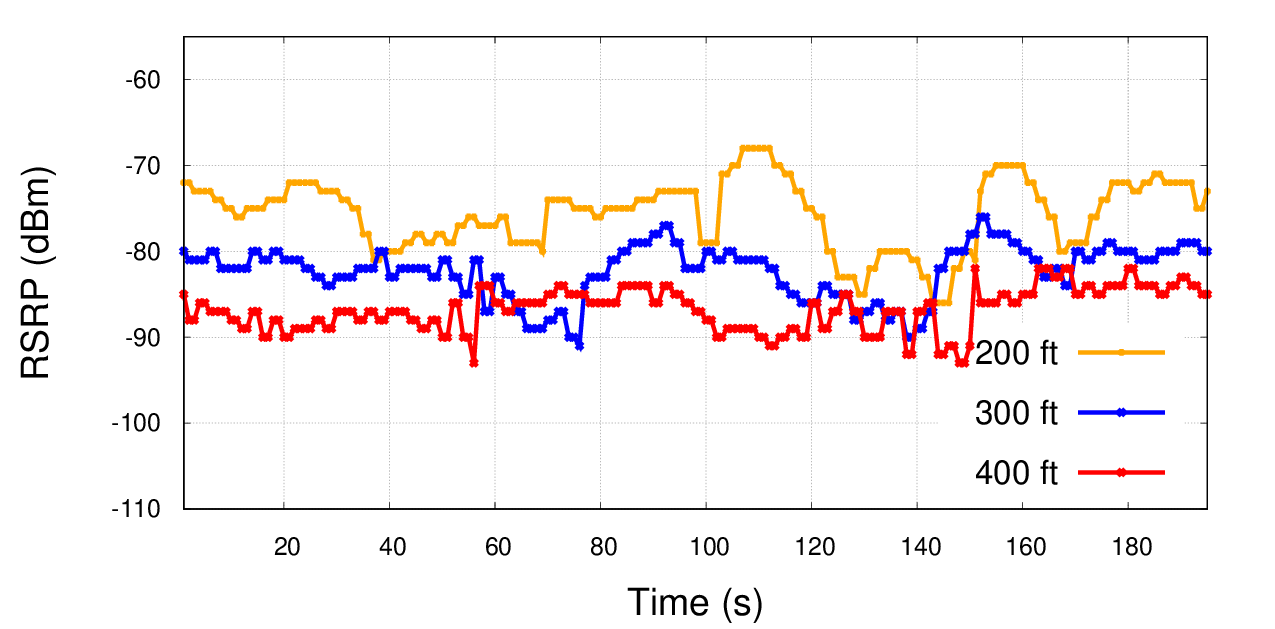}
         \caption{Mid-band 5G}
         \label{fig:rsrpMid}
     \end{subfigure}
        \caption{RSRP comparison for the combination of different elevations and network bands (speed=10 mph ) }
        \label{fig:rsrp}
\end{figure*}

\begin{figure*}
         \begin{subfigure}{0.333\textwidth}
         \centering         \includegraphics[width=\textwidth]{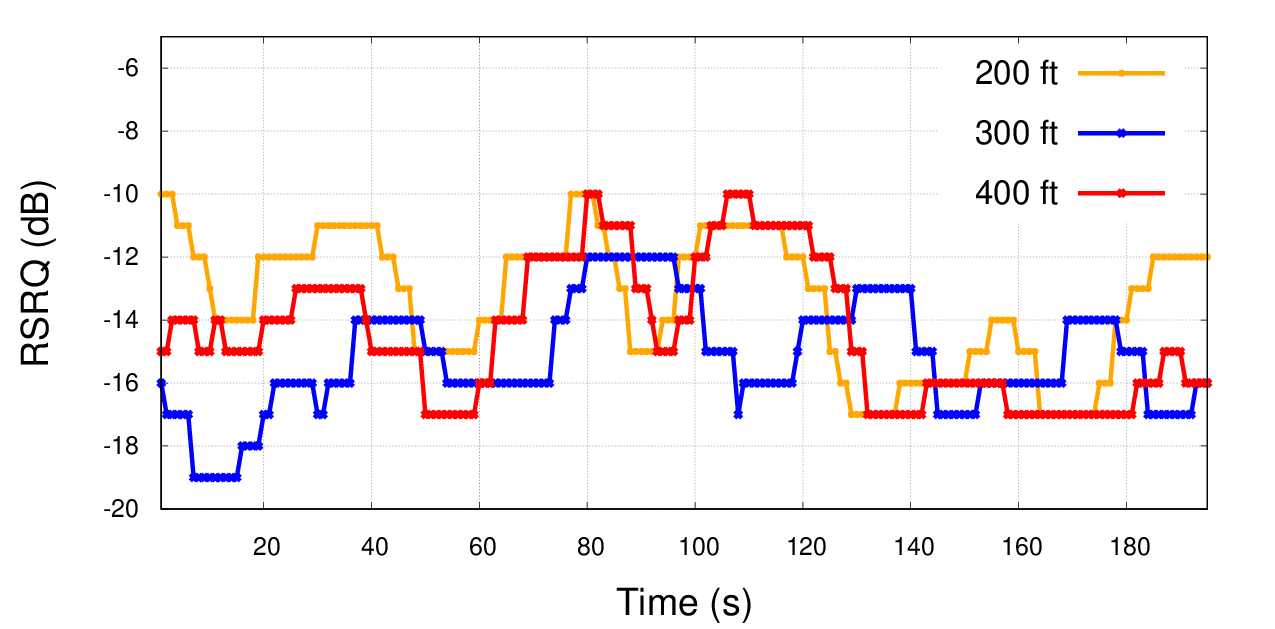}
         \caption{LTE}
         \label{fig:rsrqLte}
     \end{subfigure}
     \begin{subfigure}{0.333\textwidth}
         \centering         \includegraphics[width=\textwidth]{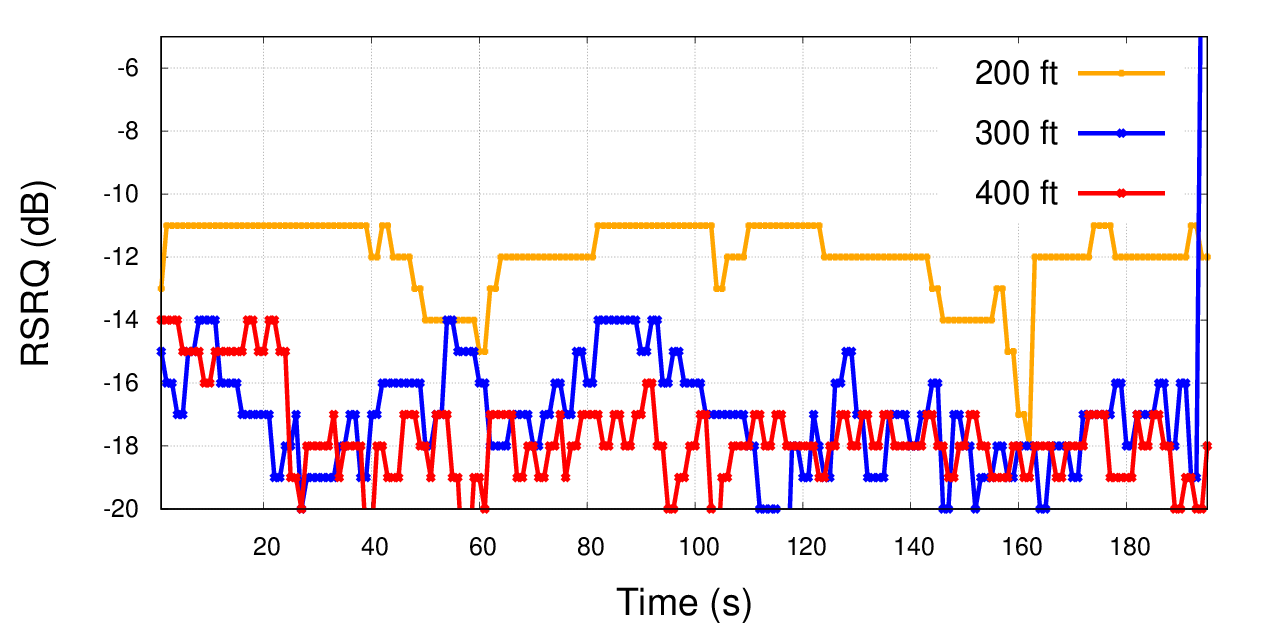}
         \caption{Low-band 5G}
         \label{fig:rsrqLow}
     \end{subfigure}
     \begin{subfigure}{0.333\textwidth}
         \centering         \includegraphics[width=\textwidth]{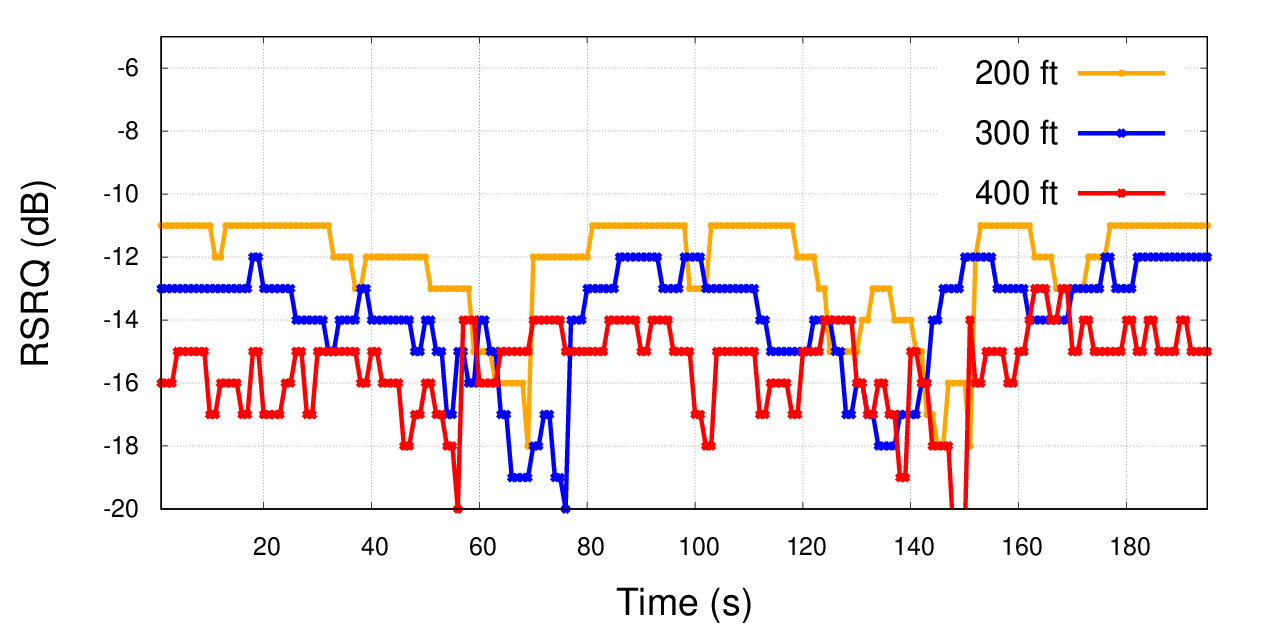}
         \caption{Mid-band 5G}
         \label{fig:rsrqMid}
     \end{subfigure}
        \caption{RSRQ comparison for the combination of different elevations and network bands (speed=10 mph ) }
        \label{fig:rsrq}
\end{figure*}

\begin{figure*}
         \begin{subfigure}{0.333\textwidth}
         \centering         \includegraphics[width=\textwidth]{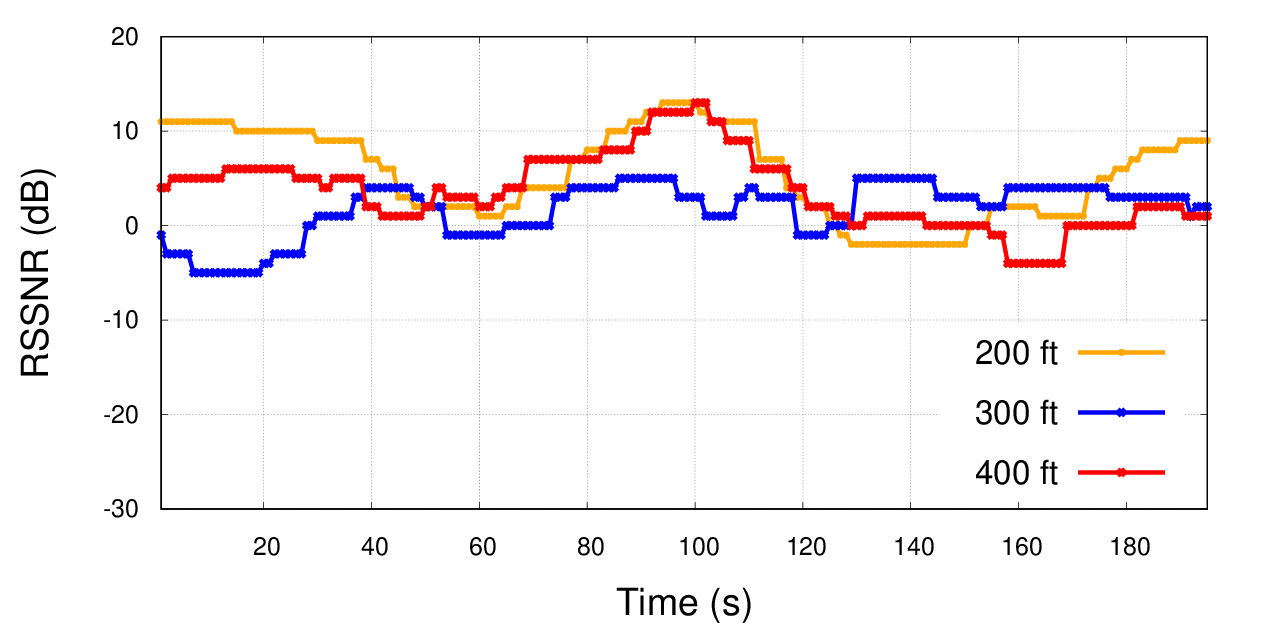}
         \caption{LTE}
         \label{fig:rssnrLte}
     \end{subfigure}
     \begin{subfigure}{0.333\textwidth}
         \centering         \includegraphics[width=\textwidth]{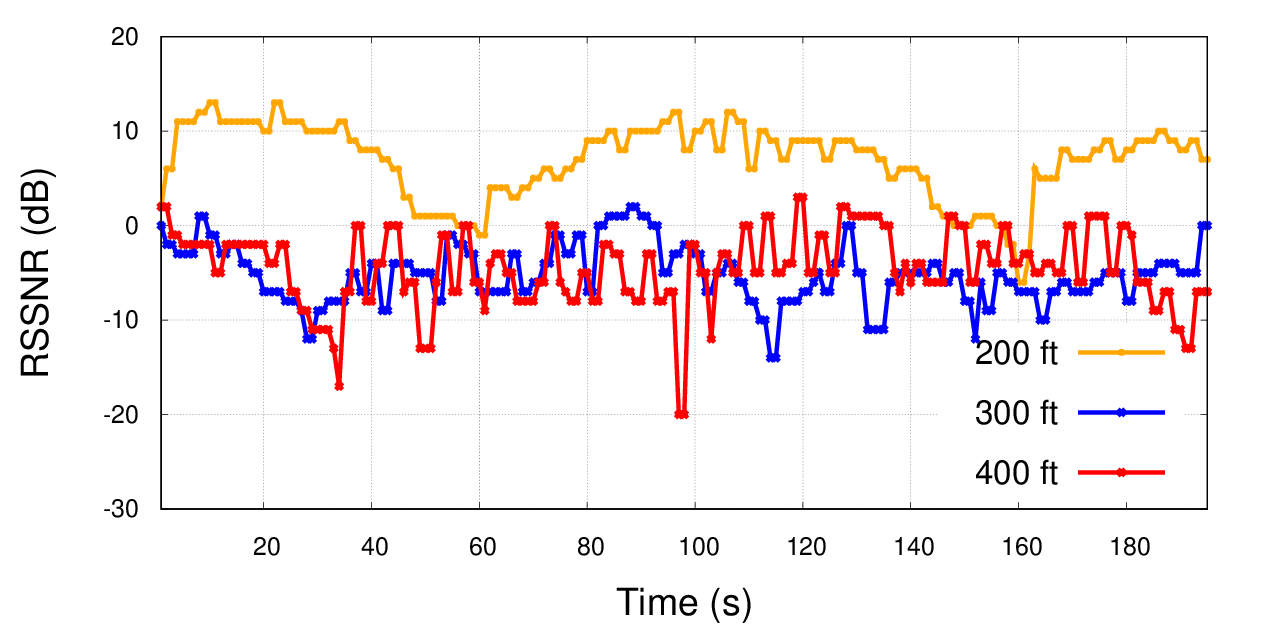}
         \caption{Low-band 5G}
         \label{fig:rssnrLow}
     \end{subfigure}
     \begin{subfigure}{0.333\textwidth}
         \centering         \includegraphics[width=\textwidth]{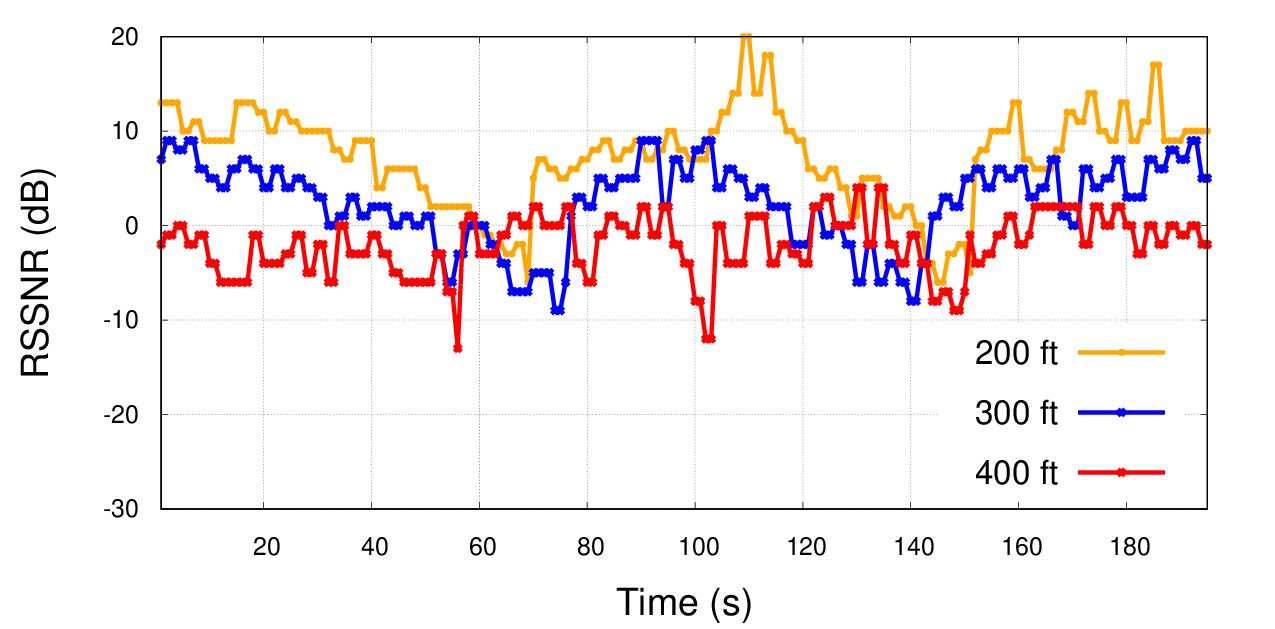}
         \caption{Mid-band 5G}
         \label{fig:rssnrMid}
     \end{subfigure}
        \caption{RSSNR comparison for the combination of different elevations and network bands (speed=10 mph )}
        \label{fig:rssnr}
\end{figure*}

\begin{figure*}
         \begin{subfigure}{0.333\textwidth}
         \centering         \includegraphics[width=\textwidth]{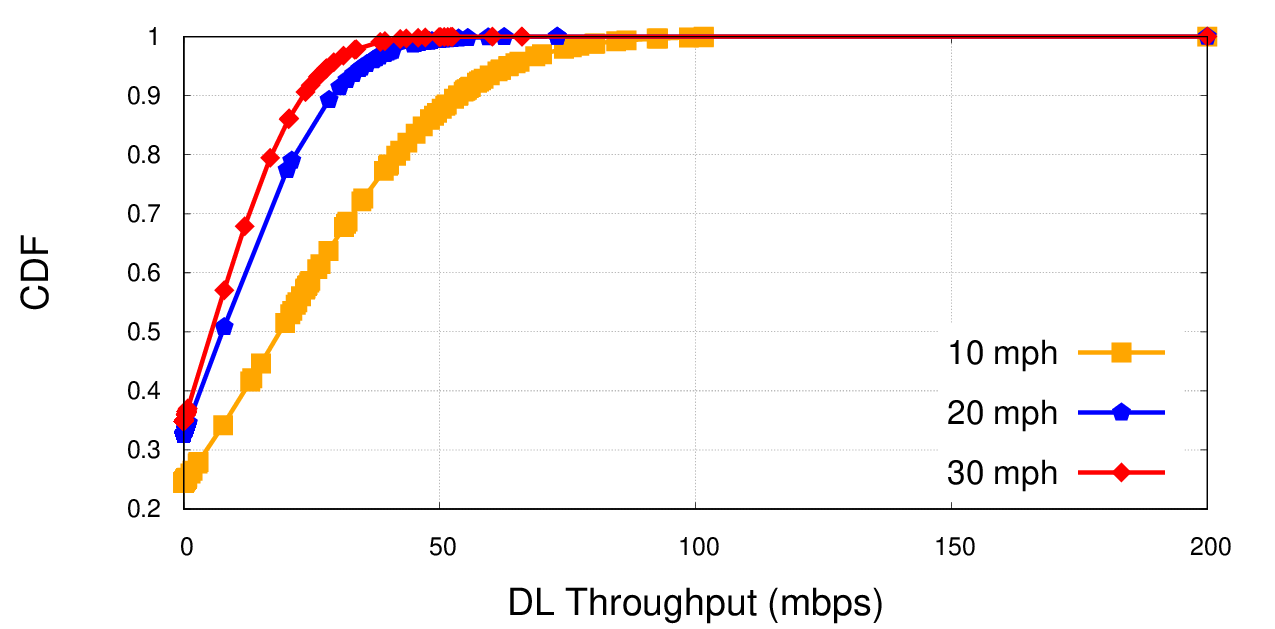}
         \caption{LTE}
         \label{fig:dlLte}
     \end{subfigure}
     \begin{subfigure}{0.333\textwidth}
         \centering         \includegraphics[width=\textwidth]{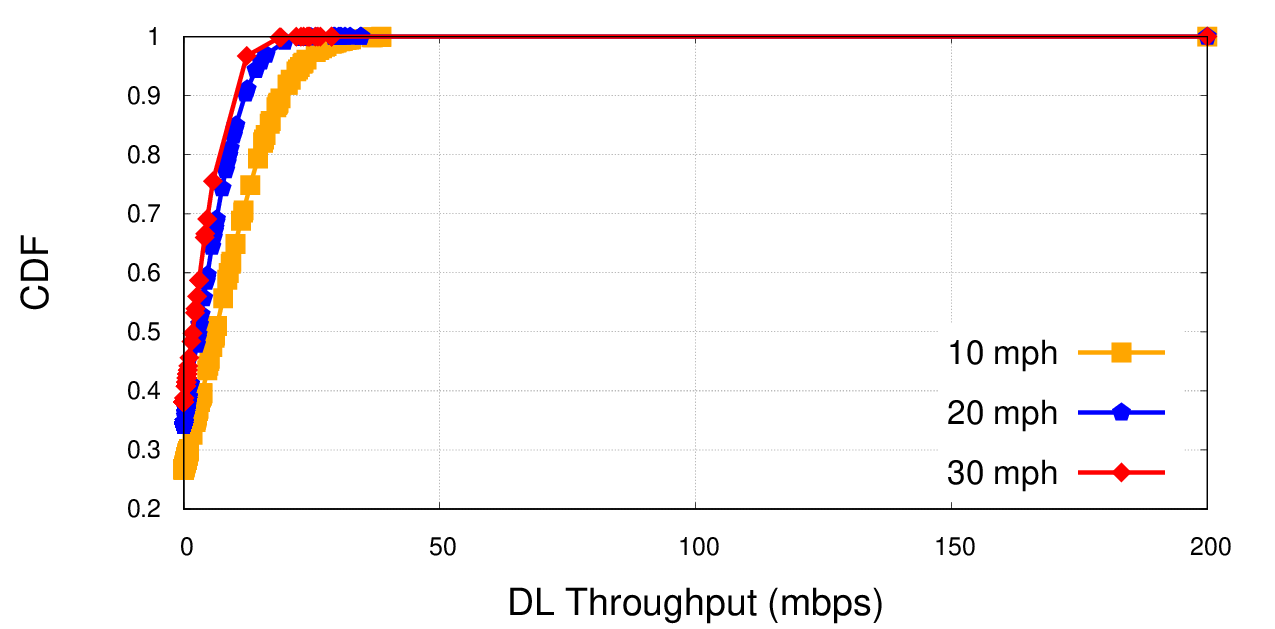}
         \caption{Low-band 5G}
         \label{fig:dlLow}
     \end{subfigure}
     \begin{subfigure}{0.333\textwidth}
         \centering         \includegraphics[width=\textwidth]{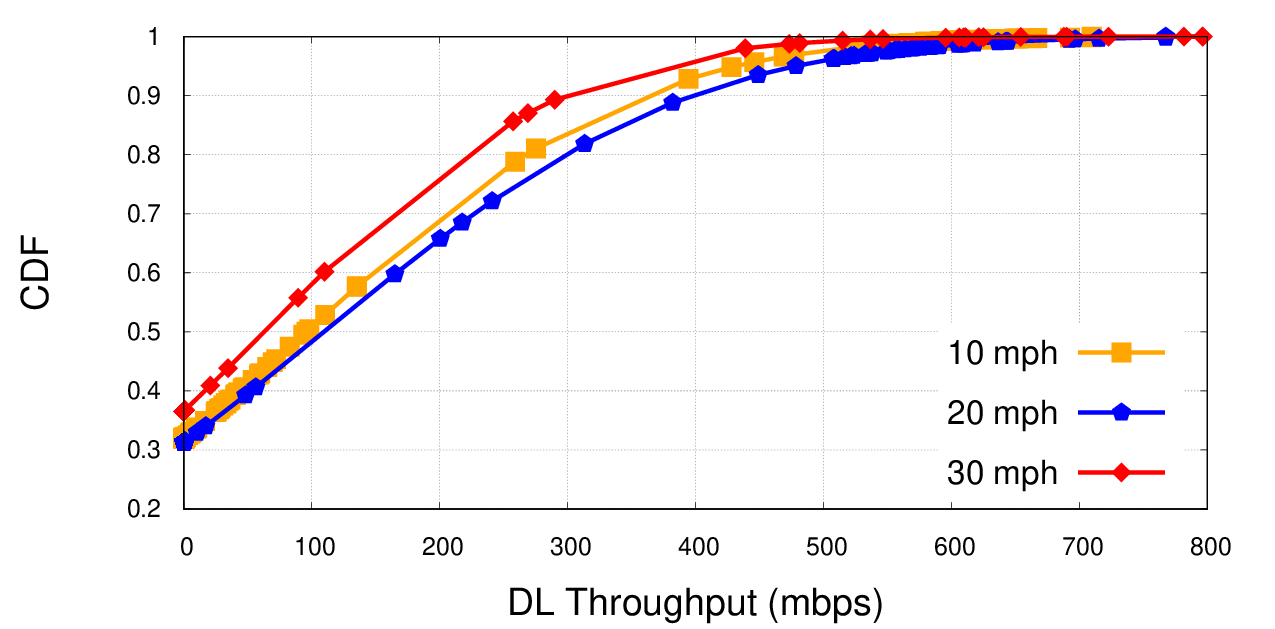}
         \caption{Mid-band 5G}
         \label{fig:dlMid}
     \end{subfigure}
        \caption{CDF of downlink throughput for the combination of different speeds and bands at the elevation of 400 ft }
        \label{fig:dlSpeed}
\end{figure*}

\begin{figure*}
         \begin{subfigure}{0.333\textwidth}
         \centering         \includegraphics[width=\textwidth]{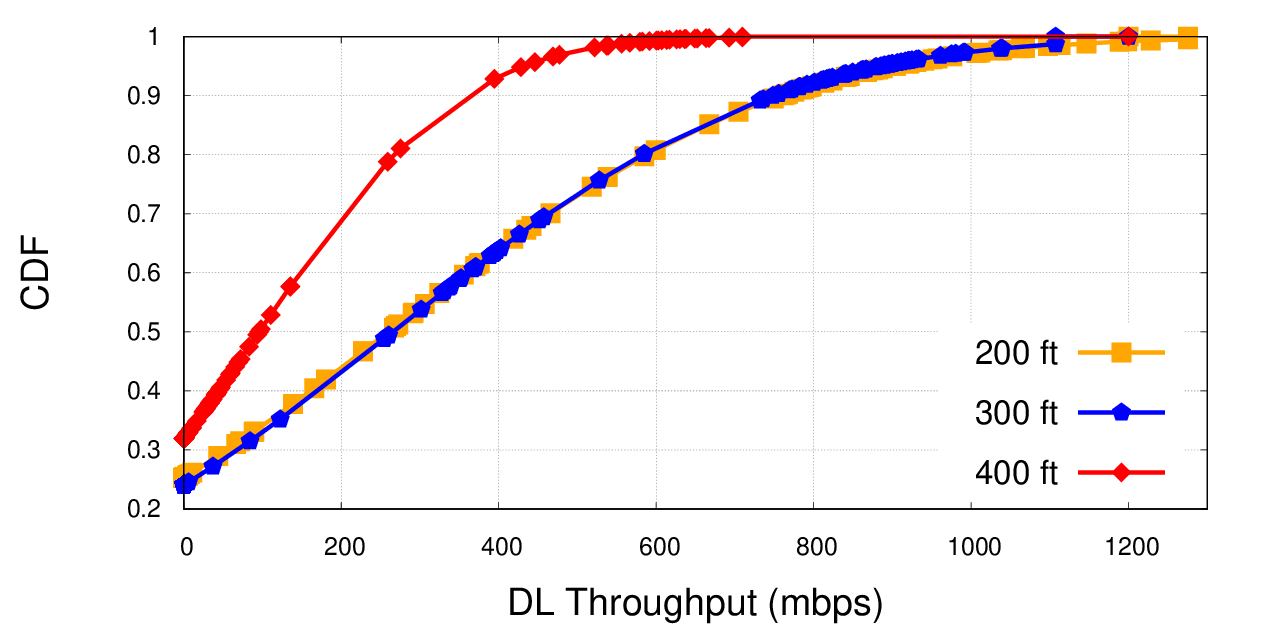}
         \caption{10 mph}
         \label{fig:dl10}
     \end{subfigure}
     \begin{subfigure}{0.333\textwidth}
         \centering         \includegraphics[width=\textwidth]{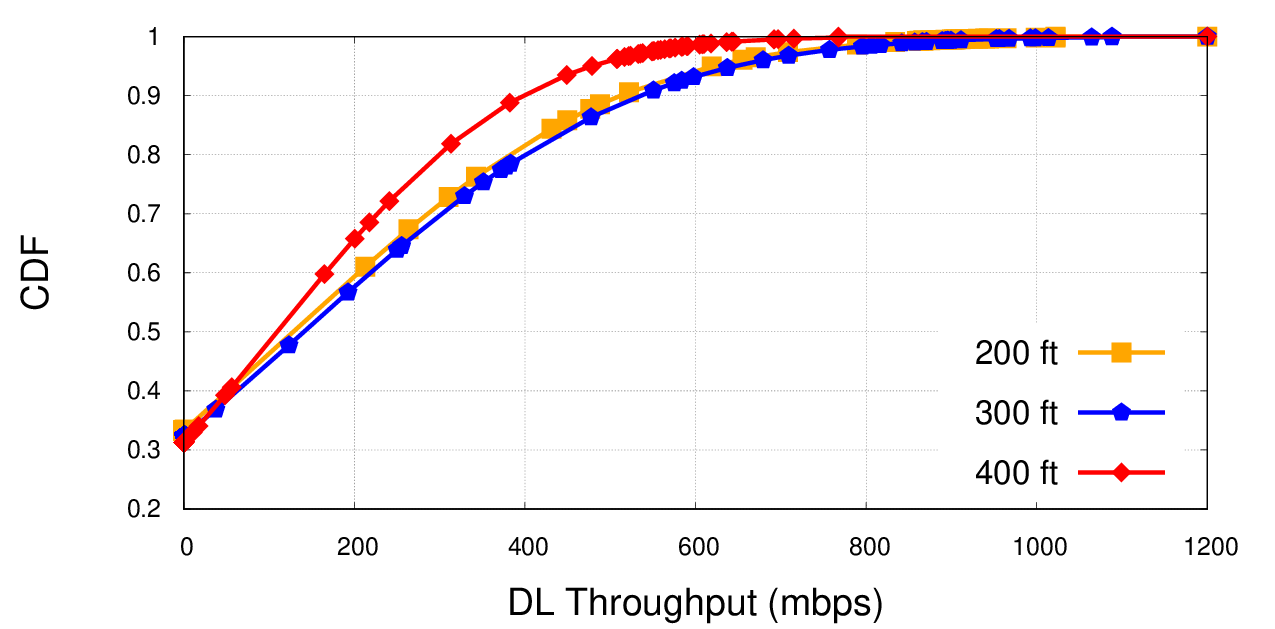}
         \caption{20 mph}
         \label{fig:dl20}
     \end{subfigure}
     \begin{subfigure}{0.333\textwidth}
         \centering         \includegraphics[width=\textwidth]{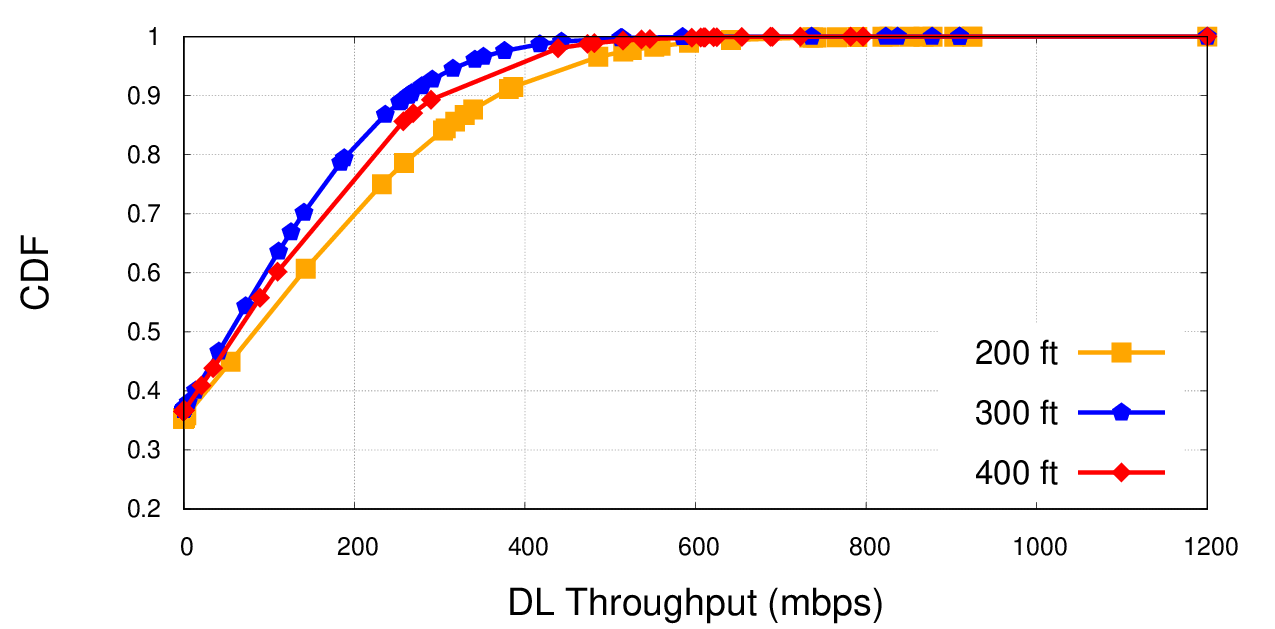}
         \caption{30 mph}
         \label{fig:dl30}
     \end{subfigure}
        \caption{CDF of downlink throughput for the combination of different elevations and speeds at mid-band 5G}
        \label{fig:dlElv}
\end{figure*}

\begin{figure*}
         \begin{subfigure}{0.333\textwidth}
         \centering
         \includegraphics[width=\textwidth]{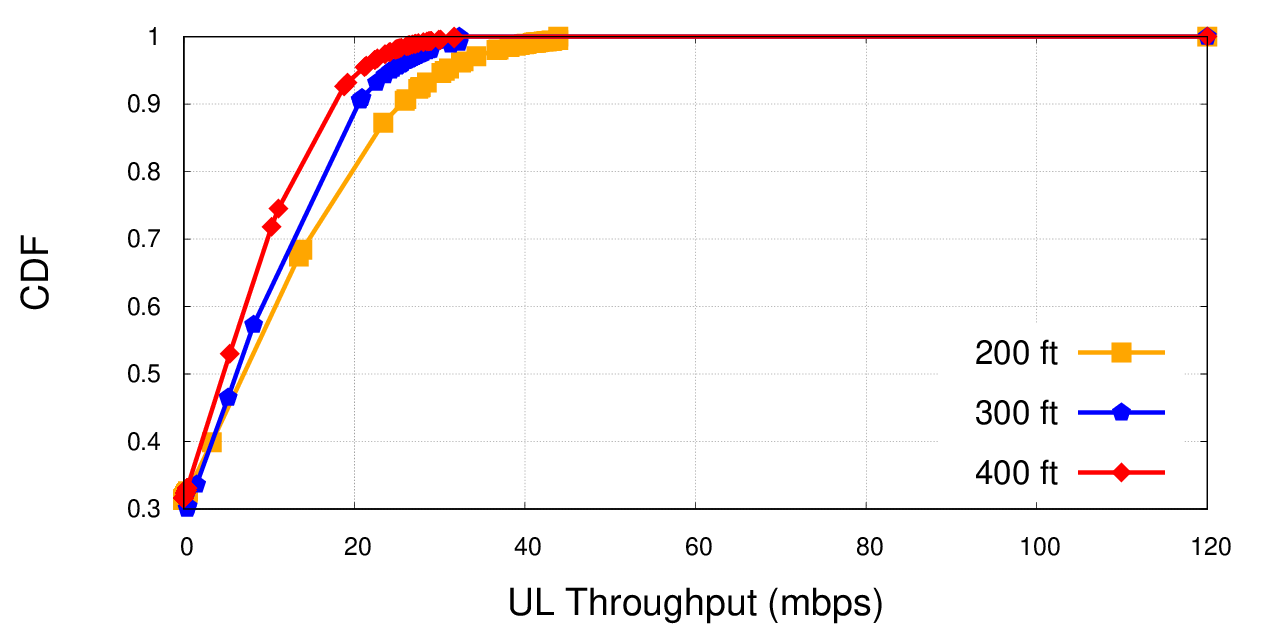}
         \caption{LTE}
         \label{fig:ulLte}
     \end{subfigure}
     \begin{subfigure}{0.333\textwidth}
         \centering         \includegraphics[width=\textwidth]{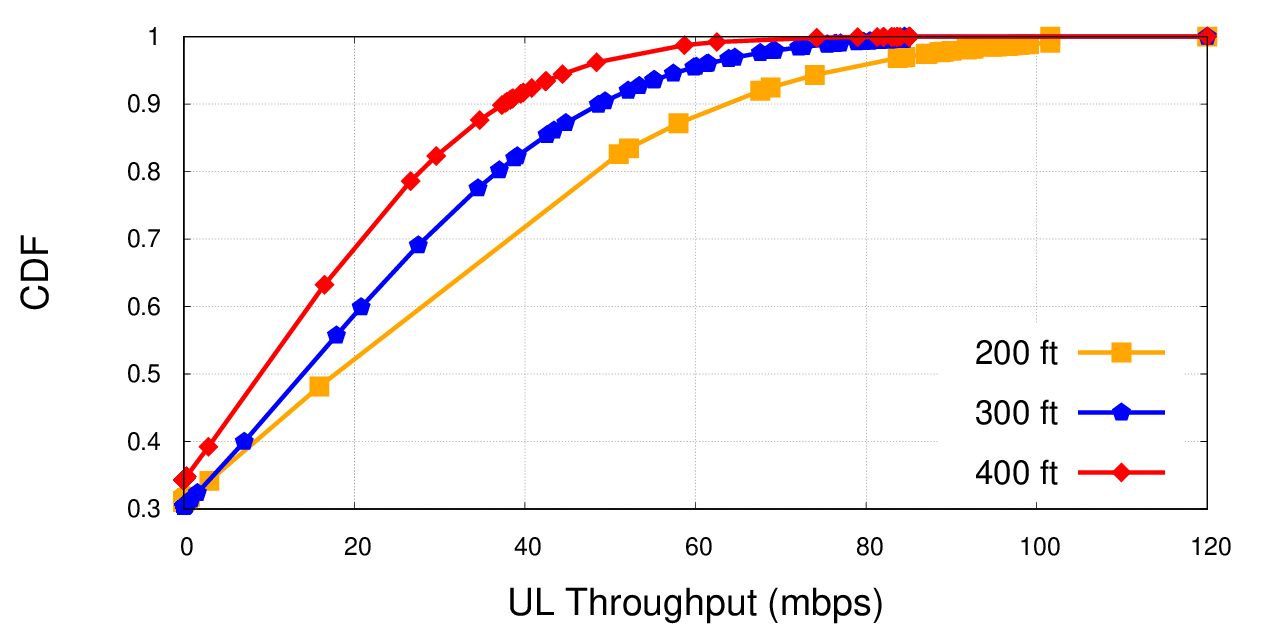}
         \caption{Low-band 5G}
         \label{fig:ulLow}
     \end{subfigure}
     \begin{subfigure}{0.333\textwidth}
         \centering         \includegraphics[width=\textwidth]{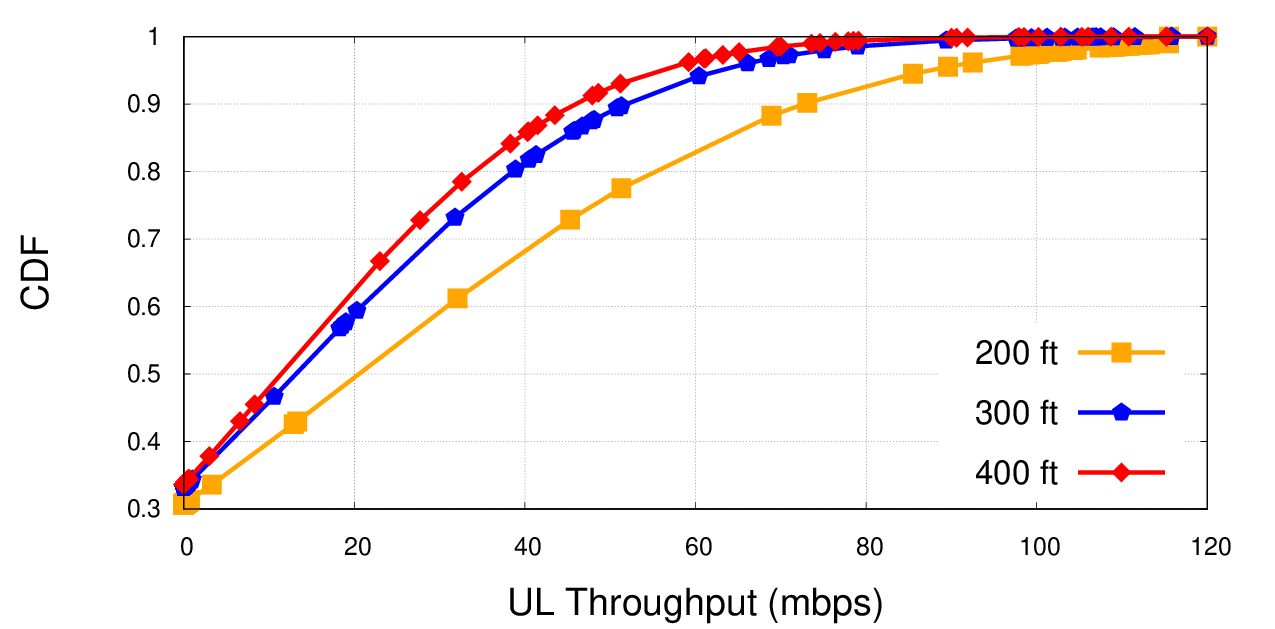}
         \caption{Mid-band 5G}
         \label{fig:ulMid}
     \end{subfigure}
        \caption{CDF of uplink throughput for the combination of different elevations and bands at the speed of 20 mph }
        \label{fig:ul}
\end{figure*}

In this section, we review the results for RSRP, RSRQ, RSSNR, downlink throughput, and uplink throughput. RSRP is the average received power from a single reference signal, measured with dBm, and ranges typically from -140 dBm for a weak signal to -44 dBm for an excellent one. Fig. (\ref{fig:rsrp}) shows the RSRP measurement for the fixed speed of 10 mph and the combination of different elevations versus different bands. We find that the low-band 5G outperforms the other bands in this metric for all the scenarios. Increasing the elevation, as the figures show, decreases the signal power in most of the scenarios. However, we find that in LTE, the 400 ft elevation outperformed the 300 ft for RSRP. It could be an indicator that, in LTE networks, increasing the elevation decreases the power, though also increases the chance of  line-of-sight signals that can increase the received power.
While due to the space limitation, it is not represented here, increasing the speed slightly decreases the RSRP. 
Fig. (\ref{fig:rsrq}) shows the measured RSRQ for the combination of different network bands versus different elevations and a fixed speed of 10 mph. RSRQ is the ratio of the signal power to the interference power, measured in dB, and ranges typically between less than -20 dB for a bad signal to more than -10 dB for an excellent signal. Just like the RSRP, increasing the elevation decreases the signal quality, in general. However, in LTE network the signal quality at 400 ft is better than that of 300 ft for the same reasoning mentioned for RSRP. Results show that the mid-band 5G has consistently higher RSRQ in comparison with low-band 5G. Mid-band 5G also outperforms LTE in the 200 ft and 300 ft settings, though LTE achieves higher RSRQ than the mid-band 5G at 400 ft. While  the uplink and downlink throughput of mid-band 5G is much higher than that of LTE, the reason behind the higher LTE RSRQ is the fact that LTE has a separated reference signal frequency for adjacent cells, while  5G generally uses the same synchronization signal block (SSB) for all cells. This may cause interference among the SSBs of adjacent cells, resulting in less signal quality or power. Our results do not indicate a significant RSRQ variation among different speeds. RSSNR measures the ratio of the reference signal power to the power of the signal noise, reported in dB. The typical range of RSSNR is starting from less than 7dB for weak signals to more than 12.5dB for excellent signals. Fig. (\ref{fig:rssnr}) compares the measured RSSNR values among different network bands for different flight elevations. The RSSNR results support the general conclusions extracted from RSRP and RSRQ figures for the same reasoning. 
 The next measured parameter is the throughput which represents the rate at which data is sent and received over the channel. From the viewpoint of the  user, the throughput is the most sensible metric for the quality of communication. Since the uplink and downlink of cellular network channels might not be symmetric, we measured the throughput for both of them separately. To measure throughput, Ookla SpeedTest was utilized, which allocates multiple simultaneous TCP connections to maximize uplink throughput or downlink throughput. Ookla operates in phases, alternating between measuring uplink and then downlink throughput for about 15 seconds each. Fig. (\ref{fig:dlSpeed}) shows the cumulative distribution function (CDF) of downlink throughput measured for the combination of different bands and speeds at 400 ft elevation. The mid-band 5G throughput is multiple times higher than that of LTE and low-band 5G, representing the high-speed download capability of this band. For better demonstration, we changed the x-axes range for LTE and low-band to [0 \quad 200] mbps. We see an obvious superiority of LTE over low-band 5G. 
 In most cases,  higher velocity decreased the downlink throughput. However, we see some exceptions as in Fig. (\ref{fig:dlMid}) the 20mph outperformed the 10mph test, which could be because of the channel condition variation due to the time difference between the two tests. Fig. (\ref{fig:dlElv}) shows a CDF comparison for the combination of different elevations versus different speeds for mid-band 5G. The 400 ft elevation test shows a significant throughput degradation compared to those of lower elevations. Last but not least, Fig. (\ref{fig:ul}) shows the CDF of uplink throughput to compare the different bands against different elevations. In this metric, the mid-band 5G represented the best performance followed by the low-band 5G.  LTE, however, did not represent comparable uplink throughput at all. The performance degradation at higher elevations is also obvious. While not reported here, the speed increment had the exact same performance degradation effect as the elevation increment in downlink throughput.

\section{Conclusion}
\label{sec:conclusion}
In this paper, we took an initial step in addressing the lack of sufficient performance evaluation studies for 5G-connected low-altitude drone communication. We exhaustively analyzed the low-band and mid-band 5G network performance and compared it to that of LTE via real measurements on a commercial cellular network in a non-urban area. We covered the impact of the elevation and velocity on the RSRP, RSRQ, RSSNR, uplink throughput, and downlink throughput. The combination of the covered KPIs, the utilized commercial network deployment, and the 3D analysis of the network performance, altogether make this work unique. We found that increasing elevation degrades network performance in almost every tested velocity-altitude combination, with negligible performance degradation observed when incrementing vehicle velocity. While the performance of the low-band 5G and LTE networks were comparable in terms of downlink throughput, we found that the mid-band 5G has manifolded throughput compared to them. For uplink throughput, we found that low-band and mid-band 5G has comparable performance much higher than that of LTE. All in all, we found the already deployed mid-band 5G is a promising candidate for aerial communication, even without any change in the network setting. %While this work is a basic step in evaluating a 5G network for aerial communications, further investigation is required to prove the efficiency of the network for aerial use. 
We planned to investigate the mmWave 5G performance for drone communication as a future direction. The study of the handover process is also another future direction.

% \section*{Acknowledgment}
% The authors would like to acknowledge the technical and financial support of T-Mobile US, Inc in providing us with access to their data collection software, providing phones for testing, and access to their network in this project.  

\nocite{*}
\bibliographystyle{IEEEtran}
\bibliography{ref}

\end{document}